# Emotion capture based on body postures and movements

*Alexis Clay\*†, Nadine Couture\*, Laurence Nigay•*

*\*LIPSI-ESTIA /† IBRLab-NTHU /•LIG-Univ Grenoble*

*Technopole Izarbel, 64210 Bidart*

*{a.clay, n.couture}@estia.fr laurence.nigay@imag.fr*

**Abstract**: In this paper we present a preliminary study for designing interactive systems that are sensible to human emotions based on the body movements. To do so, we first review the literature on the various approaches for defining and characterizing human emotions. After justifying the adopted characterization space for emotions, we then focus on the movement characteristics that must be captured by the system for being able to recognize the human emotions.

# Introduction

The goal of the Human-Computer Interaction (HCI) research field is to reduce the cognitive workload of the user, and the learning process needed to use a machine. Current systems use interaction metaphors, interactivity, feedback to reduce this workload. Present research aims at mutating the HCI field into and HII – Human Information Interaction - field: instead of working on an interface between man and machine, we want to attain a level where machines are invisible, and where human is directly interfaced to information.

Emotions are a field of growing importance in computer science. Reviewing today's research on emotion in computer science – or more restrictively in the HCI field – leads to the conclusion that most of the researches are made toward the output part of the system. Many researches are conducted to give the computer the ability to show emotions. In the field of embodied conversational agents for example, systems have been developed to allow the avatar – the virtual character displayed on the screen – to show different emotions, or to speak with a certain tone.

A smaller, though active community dedicates itself to the input part and tries to allow a computer system to understand the emotions that a user could display. The recent works of Claude Frasson [Razek, Frason 2006] on tutoring systems, for example, show a trend to imply the user's emotions (predicted or measured) in the system's reactions.

We place ourselves in this scope of research: to allow a computer system to recognize a person's emotion. Most research is made in this field on face and voice. We try to recognize a person's emotion through her gesture and body positions. The body is not the primary mean of emotional communication but still carries information about one's affective state. Such an approach can be complementary to emotion capture based on face expressions.

As an input, emotion is usually studied to adapt a system's output. Our primary aim is to amplify this emotion in an artistic context – ballet dance – and to magnify it by augmenting the stage, e.g. by displaying a visual ambience based on colours.

Nevertheless, we also aim at proposing a generic approach with a software platform that could exploit the recognized emotion for other application, including systems which goal is to modulate the interaction according to the extracted emotion.

In this paper we will present the process that we underwent to be able to create a system able to recognize emotions through gesture. It takes the form of a state of the art analysis that leads to the material we use to develop our application. At first part we will discuss about emotion: the definition of emotion, different theories about it and how to classify it. In a second time we will review some points about gesture and body positions related to emotions, before concluding.

# Emotions

*A dictionary definition*

For most of the people, the word « emotion » is one of those word that we use every day, that everybody understands, that is part of many expressions: "This painting conveys so much emotion.", "Her voice was full of emotion"… Taking the oxford online dictionary, we find that emotion is " *1 a strong feeling, such as joy or anger. 2 instinctive feeling as distinguished from reasoning or knowledge* » [AskOxford]. So, according to the context, "emotion" has two meanings [Cowie 2000]:

In the first one, the word "emotion" is taken as an entity, or a state: joy, anger are two distinct emotions. When we use this meaning we implicitly represent a discrete set of distinct emotions. This is the most intuitive way to categorize emotion and has been used by many.

In the second one, we use the word "emotion" as a generic term. In the AskOxford definition, it is put on the same level as reasoning or knowledge. This is not gratuitous. Knowledge is the raw material from which we reason, and reasoning is a cognitive process that allows us to make decisions. Emotions strongly affect memory, and thus knowledge [S&V HS 2005, p120]. Emotions allow us take decisions.

*A definition of emotion from a psychology point of view*

Taking the definition of emotion from a psychology dictionary [Grand Dictionnaire de la Psychologie 1999], we read that an emotion is:

"*A constellation of high-intensity responses which involve typical expressive, physiological and subjective manifestations*".

We can see that for psychologist an emotion is a response to a stimulus. The kind of stimulus is not defined because there is no precise stimulus that we can assess to trigger – or not trigger - an emotion. This is because the many theories on emotion are still debated: is emotion just a physiological response to the environment? Does it come from a cognitive process? Are they learnt through socialization?
People still don't agree on a general theory about where emotions come from, but the effects are pretty much defined. First, there are expressive manifestations. This is what can be perceived from the outside, and regroups facial expressions, body language, voice...

etc. The physiological expressions cannot, or with much difficulty, be repressed: those are blushing, increased heart rate, etc. And secondly there are of course some subjective manifestations that involve cognitive processes.

## *A definition drawn from psychology for computer sciences*

HUMAINE is a network of excellence which aims to create a ground research for development concerning computers and emotion [HUMAINE]. This includes capture, and analysis, which is our topic; modelling, and rendering of emotions in order to understand and act according to human emotional processes. Creating "emotion-oriented systems" is a key axis of future research in HCI, as we rely on emotion all the time. A user-centred system necessarily goes through an emotion-oriented system.

The HUMAINE network was created in January 2004 and many articles and deliverables have been published. In order to create a common vocabulary, precise enough to be used in computer sciences in an efficient way, they published [Scherer D3c 2004] a definition of emotion:

### *Emotion*

*Scherer's definition of emotion is the following: "Emotions are – "episodes of massive, synchronized recruitment of mental and somatic resources allowing to adapt to or cope with a stimulus event subjectively appraised as being highly pertinent to the needs, goals, and values of the individuals". [...]*

The following table (Table 1) is part of [Scherer D3c 2004]:

| Design Features / Types of Affect | Intensity | Duration | Synchro-nization | Event focus | Appraisal elicitation | Rapidity of change | Behavior impact |
|---|---|---|---|---|---|---|---|
| **Emotions:** angry, sad, joyful, fearful, ashamed, proud, elated, desperate | ● | · | ● | ● | ● | ● | ● |
| **Moods:** cheerful, gloomy, irritable, listless, depressed, buoyant | ● | ● | · | · | · | ● | · |
| **Interpersonal stances:** distant, cold, warm, supportive, contemptuous | ● | ● | · | ● | · | ● | ● |
| **Preferences/Attitudes:** liking, loving, hating, valuing, desiring | ● | ● | · | · | · | · | ● |
| **Affect dispositions:** nervous, anxious, reckless, morose, hostile | · | ● | · | · | · | · | ● |

*Table 1 : Characteristics of types of Affect* [Scherer D3c 2004]

As we can see, the HUMAINE network of excellence use generic concept of "affective states" which can be subdivided into five categories: emotions, moods, interpersonal stances, preferences/attitudes, and affect dispositions.

According to the table, an emotion is characterized by:

- A high level of intensity

- A focus on the event: an emotion is caused by an environmental stimulus.

- Appraisal elicitation: a cognitive process processes the triggering event.
- A high level of synchronization: the whole body tends to react according to the emotion: facial signals, posture, physiological reactions
- A great impact on behaviour
- A small duration and a high rapidity of change: emotions are a quick body response, that won't last long (but can switch category to become a mood, an interpersonal stance...), and can change quickly.

## Categories of emotion

The first interesting point comes with Darwin [Darwin: theory of Emotion 1997]. Darwin stated that an individual's emotions were based on "hereditary, biological and psychological equipment", at least the two first of those coming from evolution. His conclusion was that, much like body mutations, emotions appeared to better respond an environmental stimulus during the evolutionary process. *Emotional responses were acquired and ultimately became innate genetic characteristics*".

This theory is fairly accepted for a small set of emotions. Different theories arose in the past years.

According to Damasio [Picard 1995], an American neurologist, we are programmed at the beginning of our life to have an emotional reaction to some external, environmental stimuli, and to internal, bodily stimuli. Those are automatic reactions inherited from evolution. They participate in survival. The reaction can be split into three steps:

1. The first one is the perception of the stimulus. It can be of any kind: visual (great height, aggressive face), auditory (a roar, a rustling noise), tactile (sensation of something moving on your leg). Some are more complex and more precise: for example we are more afraid when we suddenly see a snake than a rabbit.
2. Then the brain (the amygdale) triggers a bodily reaction depending on the perceived stimulus. Hormone secretion, modification of heart beat, better irrigation of muscles.
3. We then feel the emotion (for example fear) in relation with the stimulus (aggressive face).

We are conscious of the emotion and of the stimulus, and we assume the link between those. With this association, and our personal experience, we are able to modulate the expression of those emotions throughout the course of our life.

## *A discrete set of emotion*

The first, most intuitive way to categorize emotion is to use a discrete set of labelled emotions. Each emotion – or even affective state - is represented by a word: "hate", "love", "anger"... etc. Descartes separated [Picard 1995] emotions into two sets: the primary emotions, that are pure, genuine, and the secondary emotions, which he thought to be a blending of several primary emotions. This characterization was much debated, especially about the mixed nature of secondary emotions.

The set of emotion is now still divided into two parts: the first-order emotions and the second-order emotions. The first-order emotions are the pure emotions that Descartes classified as "primary". Paul Ekman, an American psychologist identified them as a product of evolution. A basic emotion must be expressed by typical physiological changes:

1. It must be linked to universal triggering events;
2. It must appear spontaneously;
3. It must appear quickly and not last long;
4. It must be automatically appraised;
5. It must trigger some specific thoughts or sensations;
6. It must be present in other primates than human.

As we can see this list tends to somehow correlate the definition of an emotion that we retained in the first part of this paper. The 1$^{st}$ and 4$^{th}$ criteria relate to the concept of synchronization. The 3$^{rd}$ and 4$^{th}$ points are related to duration and rapidity of change. The fifth refers to the criteria of "Appraisal elicitation" in HUMAINE definition.

From these criteria Ekman did retain 6 fundamental, or basic, emotions, which can be declined into families depending on intensity:

- <u>Anger</u>: hate, rage;
- <u>Sadness</u>: sorrow, melancholy, despair
- <u>Fear</u>: anxiety, terror, panic, timidity
- <u>Joy</u>: pleasure, euphoria, ecstasy, love, devotion
- <u>Surprise</u>: astonishment
- <u>Disgust</u>: aversion.

He added in 1998 two other emotion families:
- <u>Shame</u>: guilt, humiliation
- <u>Contempt</u>

Those emotions come from evolution and are innate, as well as their corresponding reactions. They all have been observed on babies during the first months after birth. The generic property – added to the fact that psychologist agree with this theory - of these emotions makes them very suitable for a first testing set.

Throughout this subpart we hence pruned the total discrete set of emotions to Ekman's six basic emotions. In the scope of recognizing emotions through gesture, the evolutionary characteristic of these basic emotions is important as gestures and body positions are not the primary mean of emotional communication.

It is interesting to note though that a discrete set of emotion has its limitation. Those are more or less avoided when considering only Ekman's basic emotions but might appear when trying to extend the studied set of emotions.

First, this set of emotions is a mere list of labels with no real link between them. At best, emotions can be grouped into families (like "anger", "hate", "rage"). The set is not a dimensional space and doesn't have algebra: every emotion, or at least family of emotion, must be studied and recognized independently according to movement characteristics.

Second, the labels are given in a particular language and not formally described (formalisation of an emotion seems difficult if not impossible). The meanings of the six basic emotions are about the same all over the world and so the different translation may not really carry a difference of meaning. But, for other emotions, emotions translation may be a problem [Smith, Schuuichirou 2003]. For example, the word for melancholy in Japanese might not describe exactly the same affective state as the English word "melancholy".

*A continuous set of emotions*

To overcome the problems cited above, some researchers, such as Plutchik [Picard 1995], prefer to consider emotions as a continuous, multidimensional space. This space can be collapsed to a two-dimensional space. Those two dimensions are valence and arousal.

- **Valence** describes the positive/negative aspect of an emotion. Joy is clearly a positive emotion; anger, a negative one.
- **Arousal** describes the activation level of the emotion. It has been shown [Karpouzis *et al*. 2003] that emotions involve more or less dispositions to act in a certain way. Using this as a dimension, coupled with the valence dimension, allows the creation of a 2D space of emotion. For example, Fear and surprise have a high arousal level. Sadness, on the contrary, has a low arousal level.

Yet, some care is needed when handling this kind of representation. We have to keep in mind that a two-dimensional space is obtained by collapsing another high-dimensional space. There is some loss of information, and moreover the result depends largely on according to which dimensions the collapse was made. For example, Plutchik and Whissel [Picard 1995] both gave a 2D-representation of the emotion space. In Plutchik's emotion wheel fear and anger are at opposite extreme, whereas they are close together in Whissel's activation/emotion space.

In this part we have seen different means to characterize and categorize human emotions. In the scope of our project we retained Ekman's six basic emotions: **joy, anger, sadness, fear, surprise and disgust.** We chose those basic emotions according to several criteria:

They come from evolution and thus, in their purest form, they are less likely to have been altered by experience and socialization. This also means that cultural differences between various subjects should not lead to differences in their manifestations.

They fit to the HUMAINE definition, which has the advantage of being an at least semi-formal and generic categorization of affective states. This definition was proposed and approved by a wide community of scientists from various fields, from psychologists to computer scientists.

# The evolutionary point of view on emotions, and associated gesture

Our review of literature on emotions allowed us to define a set of emotions on which we will work; we now want to focus on bodily movement associated with those emotions. In this part we will review some literature about gesture and body positions.

Among humans, of course, the main mean of expression that we use to convey our internal state is the face. But many other cues are perceivable: some of the physiological cues, like blushing or getting pale; and body movement and gesture.

If we consider sadness, for example, we will notice that the exterior sides of the eyebrows lower, the eyelids contract, and the lachrymal glands are activated. But the body position also plays a role: the shoulder lower, the body is bowing. This description comes from [Dictionary of Gesture 2006]:

*Signs of sadness include* drooping eyelids*;* flaccid muscles*;* hanging head*;* contracted chest*;* lowered lips, cheeks, and jaw *(*"all sink downwards from their own weight")*;* downward-drawn mouth corners*;* raised inner-ends of the eyebrows *(i.e.,* contraction of "grief muscles"*); and* remaining motionless and passive *(Darwin 1872:176-77).*

In fact, we are able to perceive all those facts on a very precise scale. We are able to say that someone is sad even if she has a blank face. In fact, the information that we perceive from all the other cues can override the information transmitted by the face, which says "I am relax".

We usually don't analyse those body gesture and position. Our perception is rather precise, and we can perceive a signal that is so light that we are not even aware of it. We just have an overall impression about the emotional state of a person without really being able to say what triggered it.

## *Society laws and training: how we do conceal emotions.*

Moreover, in our learning process and in many cultures we are taught not to display our emotions too much, or to fake some. We often fake a smile, or restrain from showing that we are angry.

According to Desmond Morris [La Clé des Gestes 1977, p112], it is possible to create a scale of credibility of communication means, from the most credible to the least credible:

- Spontaneous signals

- Legs and feet movements
- Chest movements
- Confused and non-coded movements
- Coded hand movements
- Face expressions
- Verbal expression

Depending on the state of the individual and what he wants to show, the signals can differ. For example, if we see someone who verbally states his self confidence and shows a determined face, but whose body is bowed, we should give more credit to the chest movement. This impression that the person is not so self-confident is even more evident if he is blushing.

The spontaneous signals are the most credible because even if we are aware of them, most of the time we cannot control them. It is impossible to stop blushing. We can fake a fast breathing but it needs a lot of concentration as we need to take care of all the other signals.

The legs and feet movements are credible because we usually focus on face when trying to perceive an emotion. Those limbs are far from the point of interest, and as a reaction we tend to forget them when trying to fake an emotion or a neutral state.

The chest position reflects he muscular tonus. When we are bored for example, the chest lowers. We can fake some interest by invigorating our muscles and bend the chest but we have to keep concentrating on it.

The confused and non-coded movements are typically illustrators: gesture that we make to illustrate our speech (e.g. "I caught a fish this big") [Hartman 2004]. We are not completely conscious of them but they do enter in our vision field and so we can control them.

The coded hand movements are roughly the same things as emblems (gestures that directly translate a verbal expression, like Churchill's "V" for victory) [Hartman 2004]. We don't always perform them on purpose, but we can control them rather easily. They are on the same level as facial expressions, because even if we tend to learn more how to control our face than our hands, there are some subtleties in a face expression that we can very hardly fake or refrain. For example, even if we fake a smile, we usually don't contract the muscles around the eyes, which is a contradictory signal that might show we are not, in fact, happy.

Body movements do hence convey information about the emotions that one is feeling. Body movements for the definition of emotions that we adopted carry information less easy to capture but more credible than other kind of cues (such as facial expressions). The problem is to identify which movement – from the whole body or a specific body part – can be identified as related to an emotion.

## *Characteristics of movements: what should be considered?*

We have seen that gestures and body positions were relevant cues to detect a person's emotion: comparatively with the face, body cues carry information less easy to capture but more credible. Lots of studies have been written concerning the facial expressions: the most famous is Ekman's Facial Action Coding System (FACS), which lists and quantify the possible movements of the face through the movement of interest points (eyes and lips corners, nose tip…). There is no such list for body positions and gesture. But some researchers in various fields have found several characteristics of body position and/or movement that are related to emotions.

Marco de Meijer [De Meijer 1989] set up a very simplified categorization of movement related to emotion. He extracted 7 "dimensions of movement", which could take a value among two or three:

- Trunk movement (Stretching – Bowing): Stretching is accomplished trying to expand the body upward; bowing by bowing the trunk, lowering the head and bending the knees a bit.
- Arm Movement (opening – closing): Opening is obtained by extending the arms symmetrically away from the body, forward or upward. Closing is obtained by folding the arms across the chest.
- Vertical direction (upward – downward): upward movements should be performed by standing on the toes, bending the trunk, moving head upward. Downward movements are obtained by performing the position "bowing trunk".
- Sagittal direction (Forward – Backward): forward direction is attained by moving one step forward, backward by moving one step backward.
- Force (Strong – Light): Strong movements are developed by strengthening the muscles and clenching the fists, light by relaxing and opening the hands.
- Velocity (fast – slow) fast movements are faster, slow movements are slower than normal walking pace ( ~ 1.4 m.s$^{-1}$)

Marco de Meijer then asked four dancers to perform every possible combination of these values, recorded it on videotape and then showed those video tapes to ninety subjects. Those subjects were given answer sheets to mark which emotion they had felt and to what degree. The results were then statistically analysed.

In the scope of using it to continuously analyse a subject's movement though, this study is limited by the rough discretization of every dimension.

The resulting material is a set of "canonical movements" without any expression by themselves but which, combined in the right way, may trigger an emotion to an observer. Each emotion is labelled and the weight of every dimension of movement for each emotion is determined.

Antonio Camurri, from the University of Genova, grounded his studies on emotion and dance on Laban's Theory of Effort [Camurri 2003]. The characteristics of movement that he withdraws from Laban are the following:

- Quantity of Motion: the amount of movement mass*speed measured at a certain time;
- Contraction Index: a measure of how much space does the body visually occupies (an englobing square or cube);

The Quantity of motion through time is then analysed to perform a motion segmentation that will label motion phases and pauses. This allows measuring the rythmicity of the movement through two characteristics:

- Motion fluency, which evaluates if the movement is made with frequent sops and restarts, or if it is performed in a smooth way;
- Impulsiveness, which is calculated from the acceleration of a body part (derivative of the QoM).

We retained from this literature the different characteristics identified by De Meijer and Camurri. Camurri's movement characteristics will be used "as-is", or with slight modifications due to the fact that Camurri's work is image based whereas we plan to rely on motion capture systems which allow better 3D positioning and precision. We will also use De Meijer dimensions of movement, but we will have to ignore two of them : Force, as we cannot measure muscle tension with a motion capture system; and Directness, as real-time constraints make it difficult to define what is the beginning and the end of a gesture. One key issue related to these movements to be captured is the required precision of the capture process.

## **Conclusion**

In this paper we have reviewed the characteristic spaces for emotions. We concluded this review by the definition of emotions adopted in our project. We then focused on the body movements that must be captured according to the adopted definition of emotions. Indeed the scope of our project is to create a system– currently in design – able to capture user's body positions and orientations, as well as body movement dynamics, to extract the emotion that the user expresses. Our application domain is ballet dance and our goal is to capture the emotions that the dancer is conveying and to amplify it (for example by displaying a colour background on the stage that conveys the recognized emotion. Our application domain is adequate for a first system based on emotions captured from body movements since the dancers exaggerate the movements for conveying emotions.

In addition to the emotion capture based on movements, our goal is to provide a reusable modular software platform that can be used in another application domain. Moreover we want to use the software platform for experimental studies and tuning of the system by non-computer scientists including psychologists or gesture specialists. To do so we need to define graphical interfaces for modifying the components of the platform that are based on the right concepts for psychologists or gesture specialists.